# Bayesian mechanics of self-organising systems

Takuya Isomura

Brain Intelligence Theory Unit, RIKEN Center for Brain Science, 2-1 Hirosawa, Wako, Saitama 351-0198, Japan

Corresponding author email: takuya.isomura@riken.jp

**Abstract**

Bayesian mechanics provides a framework that addresses dynamical systems that can be conceptualised as Bayesian inference. However, elucidating the requisite generative models is essential for empirical applications to realistic self-organising systems. This work shows that the Hamiltonian of generic dynamical systems constitutes a class of generative models, thus rendering their Helmholtz energy equivalent to variational free energy under the identified generative model. The self-organisation that minimises the Helmholtz energy entails matching the system's Hamiltonian with that of the environment, leading to the ensuing emergence of their generalised synchrony. In essence, these self-organising systems can be read as performing variational Bayesian inference of their interacting environment. These properties have been demonstrated using coupled oscillators, simulated and living neural networks, and quantum computers. This framework offers foundational characterisations and predictions regarding asymptotic properties of self-organising systems interacting with their environment, providing insights into potential mechanisms underlying the emergence of intelligence.

**Main**

Intelligence—including that of biological organisms—represents one of the most intricate and paramount subjects within modern physics, demanding further exploration. As in the historical success of the theory of relativity and quantum theory, extensions of physical notions play a crucial role in elucidating previously enigmatic subjects. In pursuit of characterising intelligence, the current work investigates physical systems wherein both particle trajectories and system parameters minimise a shared Helmholtz energy.

In theoretical neurobiology, the brain is posited as an organ that creates hypotheses about the external milieu in the form of a generative or world model [1–3]. The free-energy principle has emerged as a framework that accounts for perception, learning, and action in all biological organisms in terms of minimising variational free energy, as a tractable proxy to minimise the surprisal of sensory inputs [4,5]. This notion has evolved into Bayesian mechanics, suggesting that a non-equilibrium steady state of any dynamical system that interacts sparsely with the environment can be conceptualised as a Bayesian inference of the interacting external milieu [6–10]. These notions are essential for characterising intelligence in terms of statistical inferences and the principle of stationary action. However, several challenges persist with conventional formulations, including the non-self-evident assumption that physical systems inherently minimise the surprisal and the questionable applicability to non-steady-state dynamics.

These issues can be reframed from a statistical perspective. According to the complete class theorem [11–13], a system that minimises a cost function under uncertainty—in terms of admissible rules—effectively performs a Bayesian inference of unobservable variables. In light of this, our previous work established that any neural network—characterised by the minimisation of a shared cost function—inherently engages in variational Bayesian inference [14–16]. Consequently, a reverse engineering approach was formulated, culminating in the identification of a specific generative model and variational free energy corresponding to canonical neural network architectures. This approach offers the explainability of variational Bayesian inference for characterising any canonical neural network, including its dynamics and functions. Moreover, the validity of this equivalence has been substantiated by in vitro neural networks, demonstrating that their self-organisation adheres to the identified variational free energy [17]. However, prior research has operated under assumptions that the environment is expressed as a discrete state space and the internal system persists deterministic dynamics.

The current work establishes a reverse engineering scheme for stochastic dynamical systems interacting with a continuous state-space environment—and shows that these systems can be cast as variational Bayesian inference—when states and parameters jointly minimise a common Helmholtz energy. As a corollary, it reveals that these autonomous systems self-organise towards the asymptotic alignment of their Hamiltonian with that of the external environment, a phenomenon termed Hamiltonian matching. These properties are illustrated through examples from neuroscience and physics, including real-world instances from in vitro neural networks and variational quantum eigensolvers (VQEs) [18]. This notion contextualises a generic class of self-organising systems through the lens of Bayesian inference.

**Self-organising systems minimise the Helmholtz energy**

First, a generic Langevin system that interacts with the environment is examined, where $x(t)$ represents state variables and $o(t)$ denotes sensory inputs (**Fig. 1a**, left). This system generates feedback responses or actions $a(t)$. Hereafter, the notation for the function of time $(t)$ is omitted for simplicity. Applying the Helmholtz decomposition to a Langevin equation yields the following:

$$\begin{cases} \dot{x} = f(x) - \partial_x V(x, o) + \xi_x \\ \phantom{\dot{x} =} a = g(x) + \xi_a \end{cases} \qquad (1)$$

where $\dot{x} = \partial_t x$ represents the real-time derivative of $x$; $-\partial_x V$ is the gradient flow on the scalar potential $V$; $f$ is the solenoidal flow driven by a vector potential that forms an attractor persisting even after $V$ is minimised; $g$ is an arbitrary function; and $\xi_x$ and $\xi_a$ are zero-mean white Gaussian noises, respectively. The noise refers to thermal noise within the system, while all external inputs are denoted by $o$. When the unpredictable part in $o$ is small (i.e., sparse coupling), its contribution can be involved in the gradient flow; thus, $f$ is a function of only $x$ (refer to Methods section for further details).

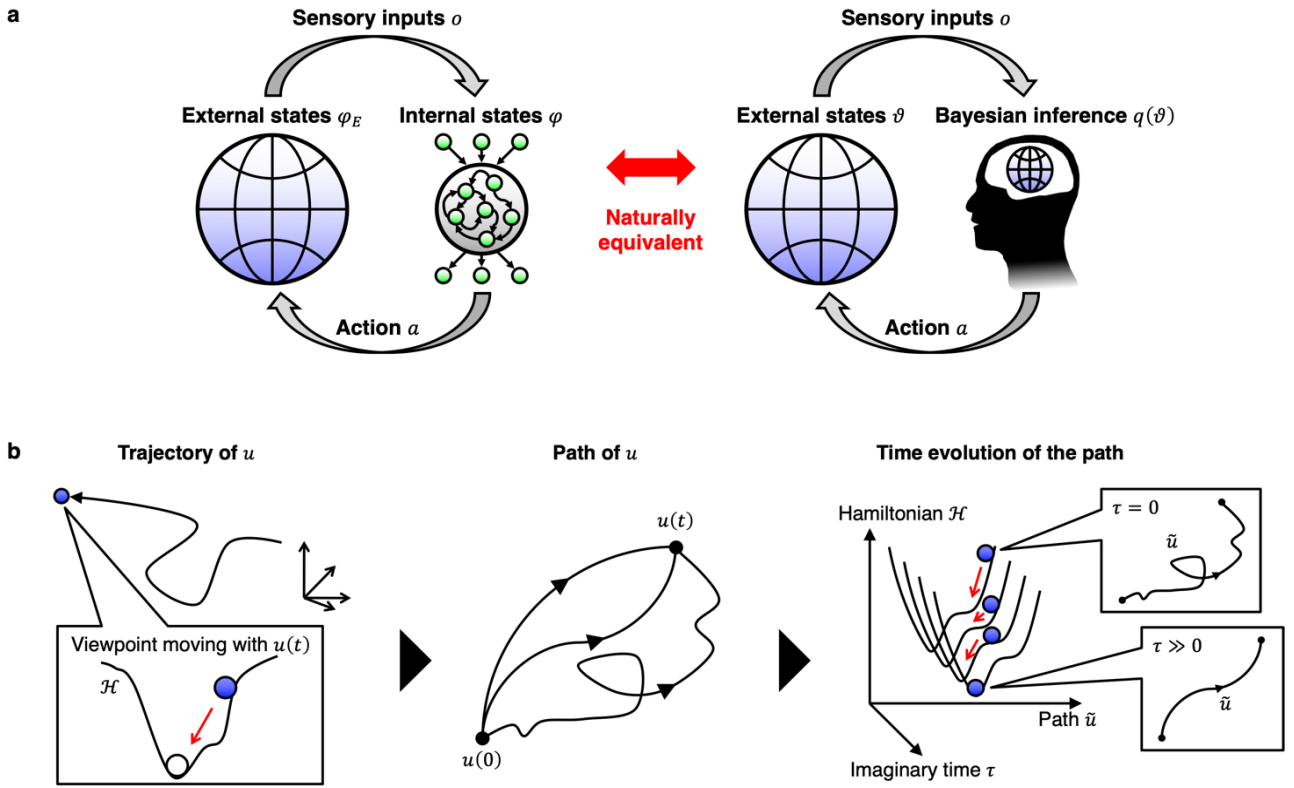

**Fig. 1. Characterising self-organising systems in terms of variational inference. a,** Characterising dynamical systems through the lens of Bayesian mechanics. When the dynamical system receives

sensory inputs *o* from the external world and generates feedback through action *a* (left), the system can be cast as variational Bayesian inference of external states (right) based on the generative model characterised by the system's Hamiltonian (refer to the main text for further details). **b,** Imaginary-time evolution of the path. Left: A view of a coordinate moving with particle $u$. In real time, the motion of particle $u$ (equation (3)) is driven by gradient flow $-\partial_u \mathcal{H}$ and noise, which continues to minimise $\mathcal{H}$ (equation (2)). Its steady state pursues a persistent attractor. Centre: Schematic of the path of $u$. Among various paths between start and end points, paths with smaller Hamiltonians are selected with higher probability, resulting in the steady-state distribution $\pi(\varphi)$. Right: The update of a path as an evolution in the imaginary time. Unlike the particle's dynamics (left), the path vector $\tilde{u}$ (right) that follows the imaginary-time gradient descent converges to the equilibrium state that minimises $\mathcal{H}$, rendering the subsequent analysis considerably more tractable. This can be read as the replacement of the principle of stationary action using the Wick rotation [19]. These distinctions hold paramount significance when employing Bayesian mechanics in contexts such as artificial and biological neural networks and quantum computations.

In the standard literature on Bayesian mechanics [6–10], $V$ is associated with surprisal. However, in those works, anti-symmetric matrix $Q(x)$ that characterises the solenoidal flow, $f = Q\partial_x V$, is selected independently of $V$, preventing $Q$ from being optimised by minimising $V$. Moreover, earlier research commonly assumed a Markov blanket, which implies conditional independency between internal and external states under a non-equilibrium steady state [6,9].

Nevertheless, the self-organising dynamics often exist in non-steady states, a discrepancy inadequately addressed by conventional treatments. Even with a recently developed path integral formulation [10], the non-steady-state dynamics have yet to be fully described.

To address these issues, here a Hamiltonian that derives both gradient and solenoidal flows is considered. By adopting a notation of (second order) generalised coordinates of motion [4,9], a new variable $x'$ is introduced to represent the change (i.e., velocity) in $x$. For simplification, a set of state variables is expressed as $u = \left(x^T, x'^T, a^T\right)^T$; the time series or 'path' of $u$ in real time from zero to $t$ is denoted as $\tilde{u} = u[0 \leq t' \leq t]$; and the internal states of the system $\varphi$ are given as a set of the path $\tilde{u}$ and system parameters $\phi$, $\varphi = \{\tilde{u}, \phi\}$. The Hamiltonian of the system is defined as a function of $\varphi$ and input sequence $\tilde{o} = o[0 \leq t' \leq t]$, as follows:

$$\mathcal{H}(\varphi, \tilde{o}) = \int_0^t \left\{ \frac{1}{2}|x' - f(x)|^2 + \frac{1}{2}|a - g(x)|^2 + \frac{\Delta t}{2}|\dot{x} - x'|^2 + V(x, o) \right\} dt' + C \quad (2)$$

where $\Delta t$ ($0 < \Delta t \ll 1$) denotes the time constant, and $f, g, V$, and the integration constant $C = C(\phi)$ are parameterised by $\phi$. Based on this, the dynamics that pursue the gradient descent on $\mathcal{H}$ (**Fig. 1b**, left) are expressed as

$$\dot{u} = \frac{1}{\Delta t}\left(-\partial_u \dot{\mathcal{H}} + \xi_u\right) \quad (3)$$

The explicit form of $\partial_u \dot{\mathcal{H}}$ is provided in the Methods section. Owing to small time constant $\Delta t$, $x'$ and $a$ rapidly converge to $f$ and $g$, respectively. In the limit of $\Delta t \to 0$, equation (3) is asymptotically reduced to equation (1). Thus, equation (3) is a natural extension of equation (1), highlighting the plausibility of equation (2) as the cost function of the considered system.

To explore non-steady dynamics, this work focuses on the path of the particles (**Fig. 1b**, centre). By coarse-graining $\tilde{u}$ with time resolution $\Delta t$, the path can be viewed as a block vector, $\tilde{u} = (u(0)^T, u(\Delta t)^T, \ldots, u(t)^T)^T$, in which the variables at a certain time point on the path are treated as distinct state variables. This transformation of the time series into Euclidean space is known as the Wick rotation [19] and has been used implicitly in related literature [14,15,20]. In light of this, the imaginary time derivative operator $\partial_\tau$ and the ensuing dynamics of $\tilde{u}$ in imaginary time $\tau$ can be considered. As equation (3) is valid at any point on the path, it can be rewritten in column vector form as $\partial_\tau \tilde{u} = -\partial_{\tilde{u}} \mathcal{H} + \xi_{\tilde{u}}$. This imaginary time gradient descent of the path converges to the fixed point that minimises $\mathcal{H}$ (**Fig. 1b**, right). This treatment replaces the non-equilibrium steady-state dynamics of the particle with the statics of the path in equilibrium, rendering the analysis considerably more tractable.

Furthermore, system parameters $\phi$ that characterise $\mathcal{H}$ are considered to have slow dynamics that minimise the shared Hamiltonian, $\partial_\tau \phi = -\partial_\phi \mathcal{H} + \xi_\phi$, which is referred to as the conjugate update rule [15]. Hence, the internal states involving path and parameters are updated as follows:

$$\partial_\tau \varphi = -\partial_\varphi \mathcal{H} + \xi_\varphi \qquad (4)$$

Equation (4) is a family of the Langevin equation and the evolution of the probability distribution of $\varphi$ can be described by the Fokker-Planck equation (a.k.a., the Kolmogorov forward equation) [21]. When each element of $\varphi$ receives an independent noise, through a certain rescaling of $\varphi$, one can consider that these variables receive the same magnitude of thermal noise $\Sigma_\varphi = \frac{2}{\beta} I$ under inverse temperature $\beta$. For non-steady-state distribution $\pi(\varphi) = p(\varphi, \tau | \tilde{o})$ corresponding to equation (4), the Fokker-Planck equation is given as follows:

$$\partial_\tau \pi = \partial_\varphi \cdot \left( \pi \partial_\varphi \left( \mathcal{H} + \frac{1}{\beta} \log \pi \right) \right) \quad (5)$$

Now, one can consider a cost function $\mathcal{A}$ that equation (5) potentially minimises. The explicit form of $\mathcal{A}$ can be reverse engineered following the previous treatment [14,15] (refer to Methods section for derivation details). Consequently, it is shown that the cost function in question has the form of the non-steady-state Helmholtz energy, which follows conventional definition [22]:

$$\mathcal{A}[\pi(\varphi), \tilde{o}] = \left\langle \mathcal{H}(\varphi, \tilde{o}) + \frac{1}{\beta} \log \pi(\varphi) \right\rangle_{\pi(\varphi)} \quad (6)$$

where $\langle \cdot \rangle_{\pi(\varphi)} = \int \cdot \pi(\varphi) d\varphi$ is the expectation over $\pi(\varphi)$. The virtue of the reverse engineering scheme is that it enables the derivation of equation (6) in a systematic and bottom-up manner without relying on heuristics or analogies. Through their construction, equations (5) and (6) provide an identical steady-state distribution—which satisfies $\partial_\tau \pi = 0$ and provides the global minimum of $\mathcal{A}$—in the form of a Boltzmann distribution:

$$\pi(\varphi) = \frac{1}{Z} e^{-\beta \mathcal{H}(\varphi, \tilde{o})} \quad (7)$$

where $Z = \int e^{-\beta \mathcal{H}(\varphi, \tilde{o})} d\varphi$ denotes the partition function.

In summary, the Hamiltonian for a class of Langevin systems was reverse engineered, and these systems were shown to potentially minimise the non-steady-state Helmholtz energy $\mathcal{A}$ (**Fig. 2, top**). These characterisations may be sufficient to elucidate straightforward dynamics, such as the synchronisation of coupled oscillators [23] (**Fig. 3a, b**). However, the treatment of more intricate scenarios—such as learning in neurons (**Fig. 3c, d**) and in the brain (**Fig. 3e, f**)—is still non-trivial in conventional physics, which will be expounded upon in the subsequent sections.

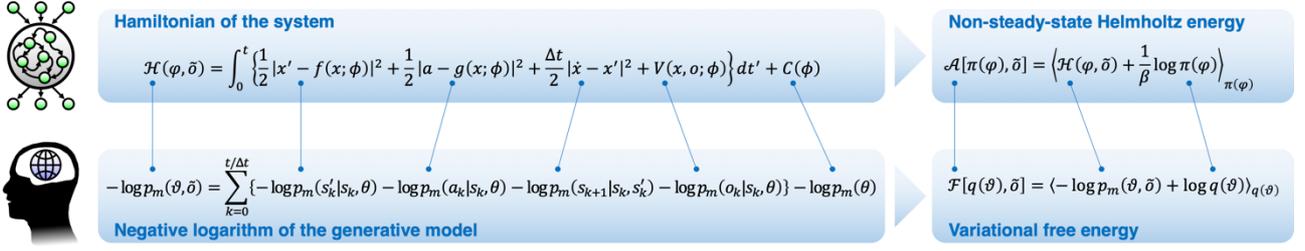

**Fig. 2. Equivalence between Helmholtz energy and variational free energy.** One-to-one correspondences of components of the Hamiltonian (top) and generative model (bottom) are shown on the left, where internal states $\varphi$ encode unobservable variables $\vartheta$. This leads to the natural equivalence between the Helmholtz energy $\mathcal{A}$ (top) and variational free energy $\mathcal{F}$ (bottom), as shown on the right. All the components of the generative model $p_m(\vartheta, \tilde{o}) = p_m(\theta) \cdot \prod_{k=0}^{t/\Delta t} p_m(s'_k|s_k, \theta)\, p_m(a_k|s_k, \theta) p_m(s_{k+1}|s_k, s'_k) p_m(o_k|s_k, \theta)$ can be characterised using the components in the original dynamical systems in equation (1). Namely, the solenoidal flow provides the conditional probability of $s'$, $p_m(s'_k|s_k, \theta) \propto e^{-\frac{1}{2}|s'_k - f(s_k;\theta)|^2}$; function $g$ determines the conditional probability of $a$, $p_m(a_k|s_k, \theta) \propto e^{-\frac{1}{2}|a_k - g(s_k;\theta)|^2}$; the subsequent state is determined based on the current state and velocity, $p_m(s_{k+1}|s_k, s'_k) \propto e^{-\frac{\Delta t}{2}\left|\frac{s_{k+1}-s_k}{\Delta t} - s'_k\right|^2}$; scholar potential $V$ gives the conditional probability of $o$, $p_m(o_k|s_k, \theta) \propto e^{-V(s_k, o_k;\theta)}$; and the integration constant characterises the prior belief about $\theta$, $p_m(\theta) \propto e^{-C(\theta)}$, where $x = s$, $x' = s'$, and $\phi = \theta$ are substituted. The generative model expresses a hypothesis about the external milieu under which a Bayesian agent operates. Thus, the internal dynamics that minimise $\mathcal{A}$ can be read as performing Bayesian inference under the generative model characterised by its own Hamiltonian.

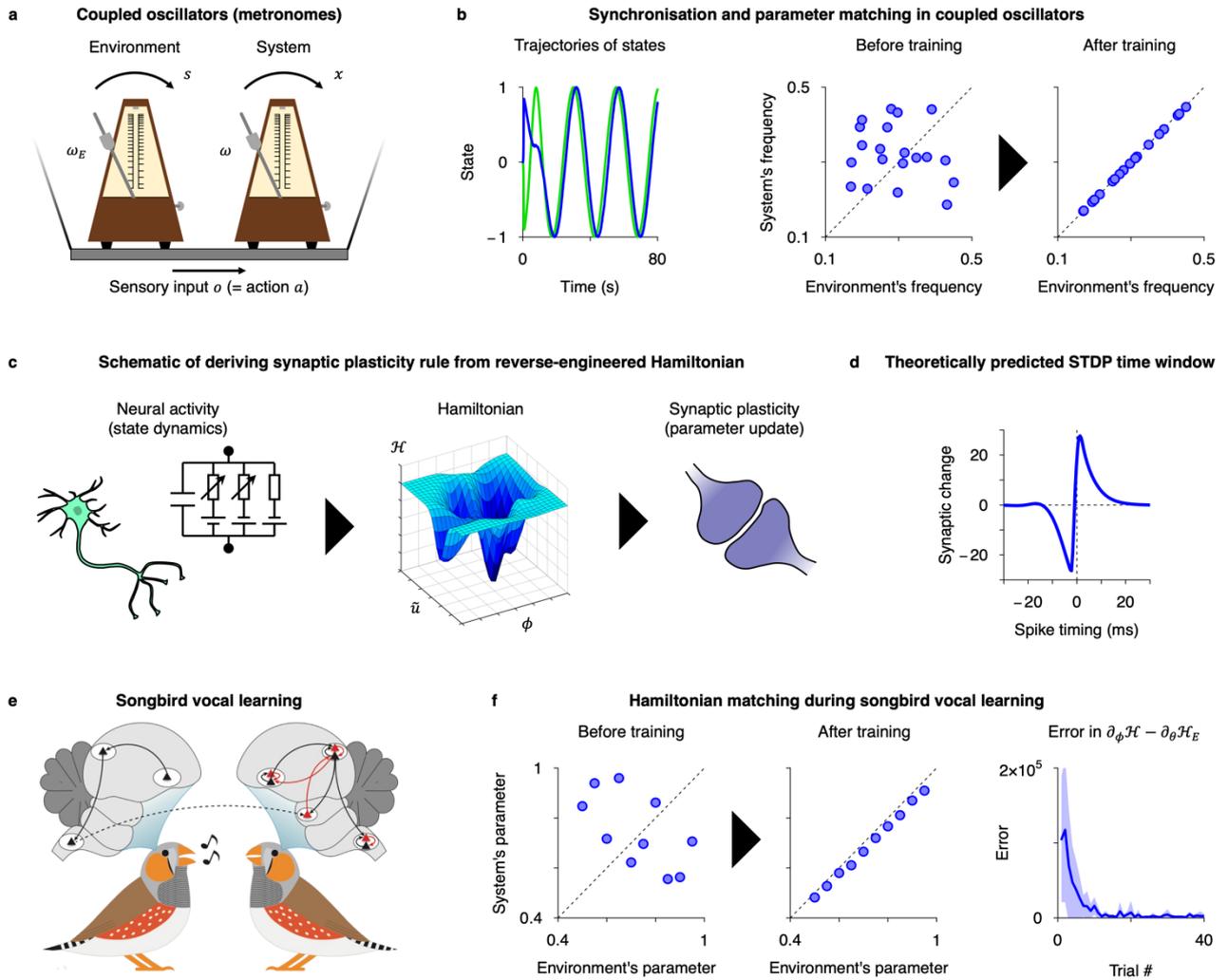

**Fig. 3. Self-organising systems minimise the Helmholtz energy. a,** Coupled oscillators (e.g., metronomes) are a simple self-organising system. The dynamics of oscillators are characterised by the Stuart-Landau equation [23] with two-dimensional states $x$ and eigenfrequency $\omega$. The system's Hamiltonian $\mathcal{H}$ is reverse engineered along with the requirements of the law of action-reaction (see Methods). Under a weak interaction with the external oscillator, they synchronise. Minimising $\mathcal{H}$ with respect to $\tilde{u}$ and $\omega$ furnishes Bayesian inference and learning (see the main text). **b,** Left: Trajectories of the system's (blue) and environmental (green) states. When the system's eigenfrequency $\omega$ is close to that of the environment $\omega_E$, frequency entrainment and phase synchronisation occur without learning. This can be viewed as a Bayesian inference of external states. However, in the absence of learning, $\omega$ differs from $\omega_E$, and the large gap

between them usually induces failure of synchronisation. Centre and right: Updating $\omega$ (by shifting the position of the mass) by following equation (6) enables successful synchronisation with the external oscillator, which corresponds to a learning of the generative model. Comparison of $\omega$ and $\omega_E$ shows that the system could learn the true eigenfrequency of the external oscillator ($n$ = 20 simulations). **c,** Reverse engineering the Hamiltonian from a realistic neuron model to derive a spike-timing-dependent plasticity (STDP) rule. **d,** Theoretically derived time window of STDP. Although it is derived from the Hamiltonian that is purely based on a neural activity model without the knowledge of STDP, its time window is highly consistent with empirically observed STDP time windows [28,29]. **e,** Generalised synchrony in songbird vocal learning. In this scenario, a juvenile or student bird learns a circuit architecture for generating a song based solely on song sequences heard from an adult or teacher bird. The circuit is characterised by two-layer Lorenz attractors [26]. **f,** Left and Centre: Convergence of the student bird's parameters with those of the teacher bird ($n$ = 10 simulations). Right: Hamiltonian matching occurring during training period. Lines and shaded areas represent mean values +/– standard deviations.

**Equivalence between dynamical systems and variational Bayes**

The virtues of Bayesian mechanics are that it provides foundational characterisations and predictions of generic dynamical system—across three distinct domains or timescales—and deploys strong analytical tools of Bayesian inference as presented below.

First, in the generic non-steady state, arbitrary dynamical systems considered in the preceding section can be cast as variational Bayesian inference (**Fig. 1a**), owing to the equivalence between the Helmholtz energy and variational free energy (**Fig. 2**). A variational Bayesian inference of unobservable variables $\vartheta$ is a process of updating the prior belief $p_m(\vartheta)$ to the corresponding (approximate) posterior belief $q(\vartheta)$ based on a sequence of observations $\tilde{o}$ [24]. Here, the environment is characterised as a process of generating sensory inputs $o$ from hidden states $s$ based on system parameters $\theta$ and system action $a$. A Bayesian agent (**Fig. 1a**, right) aims to make inferences of unobservable variables $\vartheta = \{\tilde{s}, \widetilde{s'}, \tilde{a}, \theta\}$ based on its own hypothesis about generations depicted in terms of a generative model $p_m(\tilde{o}, \vartheta)$, where the subscript $m$ depicts the model structure. This is attained by minimising the following variational free energy [4,5]:

$$\mathcal{F}[q(\vartheta), \tilde{o}] = \langle -\log p_m(\tilde{o}, \vartheta) + \log q(\vartheta) \rangle_{q(\vartheta)} \tag{8}$$

Analogous to the equivalence in the discrete state space [14–16], the non-steady-state Helmholtz energy $\mathcal{A}$ in equation (6) is formally homologous to equation (8), that is, $\mathcal{A}[\pi(\varphi), \tilde{o}] \equiv \mathcal{F}[q(\vartheta), \tilde{o}]$. This is a corollary of the complete class theorem [11–13] that states that for any pair of admissible decision rules and cost functions, there is at least a Bayesian (generative) model that renders the decisions Bayes optimal. Specifically, when $\varphi = \vartheta$ and $\beta = 1$, the Hamiltonian constitutes the generative model $p_m(\vartheta, \tilde{o}) \propto e^{-\mathcal{H}(\varphi, \tilde{o})}$ and the internal state distribution $\pi(\varphi)$ corresponds to $q(\vartheta)$ in the non-steady state, $\pi(\varphi) \equiv q(\vartheta)$. The one-to-one mapping of components of the generative model and those of the Hamiltonian is explicated in **Fig. 2** and its legend. This speaks to the natural equivalence (isomorphism) [25] of the dynamical systems defined in equation (1) and a class of variational Bayesian inference. Therefore, the process of reaching a steady state by minimising $\mathcal{A}$ can be read as the process of variational Bayesian inference that computes the posterior distribution by minimising $\mathcal{F}$. Having said this, an arbitrarily selected generative model may differ from the true generative process of the external world; thus, it must be optimised by minimising $\mathcal{A} \equiv \mathcal{F}$.

Second, in the steady state under a certain input sequence $\tilde{o}$, the distribution of the internal states $\pi(\varphi)$ represents the posterior distribution about the external states $q(\vartheta)$. Minimising $\mathcal{F}$ with respect to $q(\vartheta)$ without constraint yields the posterior distribution as follows:

$$q(\vartheta) = \frac{p_m(\tilde{o}, \vartheta)}{p_m(\tilde{o})} \tag{9}$$

where $p_m(\tilde{o}) = \int p_m(\tilde{o}, \vartheta) d\vartheta$ is the marginal distribution. This illustrates the Bayes' theorem. However, it should be noted that variational Bayesian inference is typically operated using an approximate posterior distribution, $q(\vartheta) = \prod_k q(\vartheta_k) \approx p_m(\tilde{o}, \vartheta)/p_m(\tilde{o})$, by adopting the mean-field approximation for the components of $\vartheta$ to improve analytical tractability. As a corollary of the aforementioned statement, when $\beta = 1$, equation (9) formally corresponds to the steady-state distribution in equation (7), where $\varphi$ encodes or parameterises $\vartheta$.

Finally, in the steady state over the input distribution $p(\tilde{o})$, the system's Hamiltonian aligns with the environment's Hamiltonian, a phenomenon referred to as Hamiltonian matching. The environment is characterised by a set of external states $\varphi_E = \{\tilde{s}, \tilde{s}', \tilde{o}, \theta\}$, non-steady-state distribution $\pi_E(\varphi_E)$, and Hamiltonian $\mathcal{H}_E(\varphi_E, \tilde{a})$. Subsequently, the gap in their Helmholtz energies is given by $\Delta\mathcal{A} = \langle \mathcal{A} - \mathcal{A}_E \rangle_{p(\varphi,\varphi_E)} = \left\langle \mathcal{H} + \frac{1}{\beta}\log\pi - \mathcal{H}_E - \frac{1}{\beta}\log\pi_E \right\rangle_{p(\varphi,\varphi_E)}$. The variation of $\Delta\mathcal{A}$, $\delta\Delta\mathcal{A} = \langle \delta(\mathcal{H} - \mathcal{H}_E) \rangle_{p(\varphi,\varphi_E)}$, constantly satisfies zero when $\delta\mathcal{H} = \delta\mathcal{H}_E$ holds. Because the minimisation of $\Delta\mathcal{A}$ leads to the zero variation, the following equations asymptotically hold:

$$\begin{cases} \partial_o \dot{\mathcal{H}} = \partial_o \dot{\mathcal{H}}_E \\ \partial_a \dot{\mathcal{H}} = \partial_a \dot{\mathcal{H}}_E \\ \partial_x \dot{\mathcal{H}} = \partial_s \dot{\mathcal{H}}_E \partial_x s \\ \partial_{x'} \dot{\mathcal{H}} = \partial_{s'} \dot{\mathcal{H}}_E \partial_{x'} s' \\ \partial_\phi \dot{\mathcal{H}} = \partial_\theta \dot{\mathcal{H}}_E \partial_\phi \theta \end{cases} \qquad (10)$$

In simple terms, $\partial_\varphi \mathcal{H} = \partial_\vartheta \mathcal{H}_E \partial_\varphi \vartheta$ in the large time limit. The derivation details are provided in the Methods section. Equation (10) summarises a foundational prediction of Bayesian mechanics: the Hamiltonian of the system asymptotically converges to that of the environment, up to a trivial indeterminacy about mappings from $s$ to $x$ and from $\theta$ to $\phi$. Under this condition, the scalar potential and functions $(f, g, V, C)$ constituting the system's Hamiltonian align with those constituting the environment's Hamiltonian $(f_E, g_E, V_E, C_E)$, except for a trivial indeterminacy. This is identical to the matching between the generative model and generative process, leading to generalised synchrony [26] between internal and external systems.

In summary, this equivalence enables us to seamlessly switch between the dynamical system and Bayesian inference formulations, harnessing their insights to interpret and elucidate a system's behaviour. The strength of Bayesian mechanics lies in its explainability to reframe any given dynamical system by actively engaging in Bayesian inferences of external states. Consequently, the self-organisation of various dynamical systems consistently progresses towards acquiring a Bayes-optimal model of the interacting external world. This notion is readily applicable to practical dynamical systems such as coupled oscillators to predict their asymptotic self-organisation (**Fig. 3a**), where they function as a Bayesian agent that infers hidden states and learns parameters of the external system (**Fig. 3b**). The remainder of the paper explores the application of this notion in practical cases within neuroscience and physics.

**Bayesian mechanics in neuronal systems**

Application of Bayesian mechanics to neuronal networks are essential for understanding the dynamics and functions of the brain. Neurons receive inputs via synaptic connections and generate spikes (i.e., neural activity) (**Fig. 3c**, left). The Hodgkin–Huxley equation [27]—a neuron model that exhibits realistic spiking activity—falls under the category of dynamical systems expressed by equation (1). Moreover, synaptic connections become plastic depending on the timing of the firing of pre- and post-synaptic neurons, a phenomenon known as spike-timing-dependent plasticity (STDP) [28,29]. If a presynaptic neuron fires first and a postsynaptic neuron fires immediately thereafter, long-term potentiation occurs; conversely, if a postsynaptic neuron fires first and a presynaptic neuron fires later, long-term depression occurs. However, earlier computational works have modelled neural activity [27] and STDP [30] separately. In contrast, recent research has emphasised the importance of a conjugate relationship between neurons and synapses that commonly minimise the shared Helmholtz energy [14–16].

To test this hypothesis, the Hodgkin–Huxley model accompanied by glutamatergic receptor models was reverse engineered to obtain the Hamiltonian by computing the integration of the state dynamics equations based on the above established scheme (**Fig. 3c**, centre). Herein, parameters $\phi$ correspond to a synaptic weight $W$. Subsequently, the equation for synaptic plasticity was derived as a gradient descent on the obtained Hamiltonian with respect to $W$ (**Fig. 3c**, right), allowing the prediction of the form of plasticity time window (**Fig. 3d**). This time window is purely predicted based on the neural activity equation without prior knowledge of STDP;

nevertheless, it is consistent with experimental results of STDP [29]. These observations suggest a certain biological plausibility of the view that neural activity and synaptic plasticity minimise the common energy, and show interesting consistency with the recently proposed prospective configuration [31].

Moreover, in neuroscience, equation (3) is commonly referred to as the predictive coding model and has been applied to model various neural computations, including visual [32] and auditory [33,34] perception and motor control [35]. In these models, each element of *u* is associated with the activity of an individual neuron or neural ensemble (see also equation (11) in Methods section for the explicit form). Through approximations, the original Hodgkin–Huxley equation can be reduced to a two-dimensional Hodgkin–Huxley model [36]. This class of models involves the FitzHugh–Nagumo model [37,38]—a neuron model homologous to the oscillators [23] illustrated in **Fig. 3a, b**—as well as the canonical neuron model [14,15], which is a type of firing rate model. Previous work [14–16] reverse engineered the Helmholtz energy $\mathcal{A}$ (a.k.a., the neural network cost function) for canonical neural networks and showed that the conjugate plasticity rule that minimises $\mathcal{A}$ exhibits the form of biologically plausible Hebbian plasticity [39]. Interestingly, the relationship between Hodgkin–Huxley model and STDP re-emerges in their approximate models. That is, the Helmholtz energy obtained based on canonical neuron models derives the equation for the widely recognised associative or Hebbian plasticity, further supporting the notion that neural activity and plasticity jointly minimise $\mathcal{A}$. As previously established, these neural networks effectively engage in variational Bayesian inferences of their interacting environment [14–16], resulting in generalised synchrony [26].

A seminal example of this is the songbird vocal learning, in which a young bird learns to imitate the song of adult birds [26,40] (**Fig. 3e**). This phenomenon can be modelled using predictive coding, and the minimisation of the shared Helmholtz energy facilitates the learning of parameters in the song generation circuit in young birds, enabling Hamiltonian matching (**Fig. 3f**). Moreover, such synchronisations have been employed to model synchronisation in neural activity [41], music perception [42], and humanoid robot control [43,44].

Regarding empirical demonstrations, the emergence of neural activity encoding the posterior beliefs about hidden states [45], as well as neuronal substrates associated with prior beliefs [46] and prediction errors [47], have been reported. Moreover, evidence that neural networks minimise variational free energy [48,49] has been reported. In such cases, the generative model and variational free energy can be computed empirically by reverse engineering the Hamiltonian and Helmholtz energy from neural activity data. A compelling example of a conjugate update of synaptic weights in living neural systems is the self-organising acquisition of blind source separation—the ability to infer and disentangle multiple causes of sensory inputs—within in vitro neural networks (**Fig. 4a**). The minimisation of Helmholtz energy—reverse engineered from empirical data—can accurately predict learning within in vitro neural networks by assimilating received structural input patterns [17]. During training, changes in effective synaptic connectivity *W* occurred in a manner that encoded the likelihood mapping parameters. Subsequently, it was confirmed that this self-organisation entailed Hamiltonian matching between the in vitro neural networks and the input signal generation process (**Fig. 4b**). These results highlight the plausibility of Bayesian mechanics for characterising the self-organisation of neuronal networks.

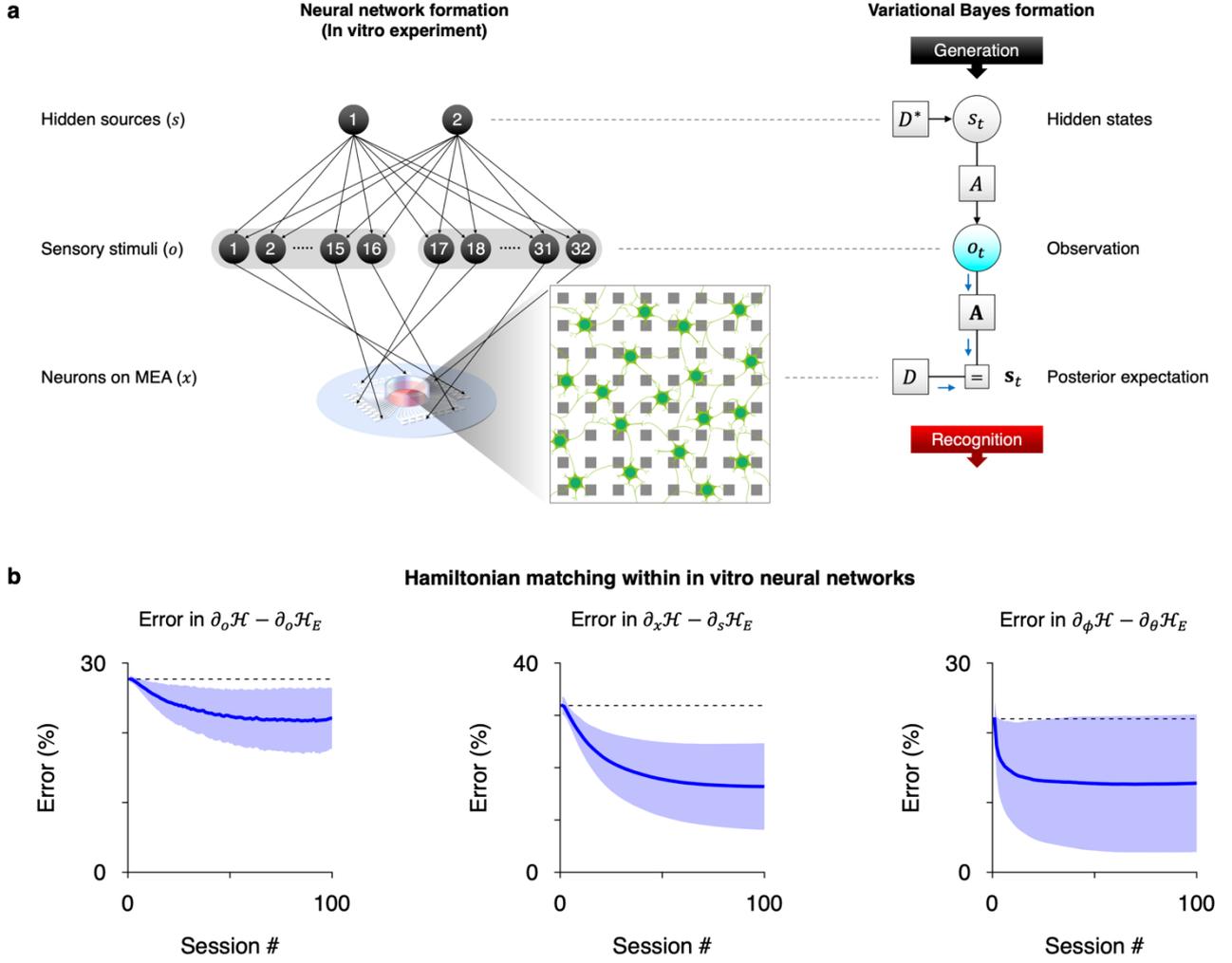

**Fig. 4. Bayesian mechanics in neuronal systems. a,** In vitro neural networks on a microelectrode array (MEA) (left) and corresponding variational Bayesian inference formulation (right). Here, 32-dimensional patterned inputs generated by mixing two-dimensional hidden states were induced to in vitro neural networks as electrical pulses. Canonical neural networks with a single feedforward layer—interacting with a discrete state-space environment—are given as $\dot{x} \propto -\text{sig}^{-1}(x) + (W_1 - W_0)o + h_1 - h_0$, where $W_1$ and $W_0$ are excitatory and inhibitory synaptic weight matrices and $h_1$ and $h_0$ are adaptive firing threshold that are functions of $W_1$ and $W_0$ [14]. These networks employ the Hamiltonian $\mathcal{H}(\varphi, \tilde{o}) = -\int_0^t \begin{pmatrix} x \\ x \end{pmatrix}^{\text{T}} \left\{ \begin{pmatrix} W_1 \\ W_0 \end{pmatrix} o + \begin{pmatrix} h_1 \\ h_0 \end{pmatrix} \right\} dt' + C$, where $\bar{x} = \vec{1} - x$ denotes the sign-flipped $x$ centred on 1/2 and $\vec{1} = (1, \ldots, 1)^T$ denotes a vector of ones. Here, one can consider that a spiking neural activity $x \in \{0,1\}^{N_x}$ is generated from a Bernoulli distribution $\pi(x) = \mathrm{Bern}(\mathbf{x})$, analogous to a Poison spiking neuron model with firing intensity $\mathbf{x}$. The dynamics of a neural ensemble forms the distribution $\pi(\varphi)$ characterised by internal states $\varphi = \{\tilde{x}, W, \phi\}$. Subsequently, the Helmholtz energy (a.k.a., neural network cost function) is defined as $\mathcal{A}[\pi(\varphi), \tilde{o}] = \int_0^t \left(\frac{\mathbf{x}}{\bar{\mathbf{x}}}\right)^T \left\{ \ln\left(\frac{\mathbf{x}}{\bar{\mathbf{x}}}\right) - \begin{pmatrix}\mathbf{W}_1\\\mathbf{W}_0\end{pmatrix} o - \begin{pmatrix}\mathbf{h}_1\\\mathbf{h}_0\end{pmatrix} \right\} dt' + C$, where variables in bold denote the ensemble averages (expectations) of the corresponding italic case variables. The gradient descent on $\mathcal{A}$ furnishes the neural activity, $\dot{\mathbf{x}} \propto -\partial_\mathbf{x} \mathcal{A}$, and synaptic plasticity, $\dot{\mathbf{W}}_l \propto -\partial_{\mathbf{W}_l} \mathcal{A}$ (for $l = 1, 0$), of canonical neural networks [14]. By fitting empirical neural activity data to $\mathcal{A}$, the generative model and variational free energy employed by the in vitro networks were reverse engineered. **b,** Hamiltonian matching occurred within in vitro neural networks during the self-organisation by assimilating sensory information. Open data accompanied with previous work [17] were analysed. Panels depict errors between the derivatives of $\mathcal{H}$ with respect to $o$, $x$, and $\phi$ and those of $\mathcal{H}_E$ with respect to $o$, $s$, and $\theta$. Lines and shaded areas represent mean values +/− standard deviations (*n* = 30 independent experiments). Please refer to previous work [17] for details regarding experiments and data analyses.

**Bayesian mechanics in quantum systems**

The applicability of Bayesian mechanics extends to quantum systems, wherein the steady-state distribution is replaced by a thermal equilibrium density matrix. For analytical tractability, this work focuses on the thermal equilibrium states and the quasi-static transitions between these states. The internal states comprise a state vector $|\psi\rangle$ and classical system parameters $\phi$, collectively denoted as $\varphi = \{|\psi\rangle, \phi\}$, where the parameters $\phi$ are treated as classical variables, akin to the treatment in usual quantum computations. In thermal equilibrium at inverse temperature $\beta$, the probability that the system employs eigenstate $|\psi_k\rangle$ and eigenenergy $E_k$ is proportional to the Boltzmann distribution $e^{-\beta E_k}$. Thus, the mixed state density matrix at equilibrium can be expressed as $\hat{\rho} = \frac{1}{Z} e^{-\beta \hat{\mathcal{H}}} = \frac{1}{Z} \sum_k e^{-\beta E_k} |\psi_k\rangle\langle\psi_k|$ using the partition function $Z = \sum_k e^{-\beta E_k}$. This $\hat{\rho}$ is associated with the probability density of the path of fast variables. Subsequently, the energy expectation $\langle \hat{\mathcal{H}} \rangle = \mathrm{tr}[\hat{\rho} \hat{\mathcal{H}}]$, von Neumann entropy $-\mathrm{tr}[\hat{\rho} \log \hat{\rho}]$, and steady-state Helmholtz energy $\mathcal{A} = -\frac{1}{\beta} \log Z = \mathrm{tr}\left[\hat{\rho}\left(\hat{\mathcal{H}} + \frac{1}{\beta} \log \hat{\rho}\right)\right]$ are expressed based on $\hat{\rho}$ and $Z$.

Similar to classical systems, updating the internal states—by jointly minimising $\mathcal{A}$ with respect to both the state vector and parameters—can be interpreted as surprisal minimisation and Bayesian inference. The thermal equilibrium density matrix $\hat{\rho}$ can be construed as encoding a mixture posterior belief that corresponds to a mixture generative model [40] in which the probability of selecting a generative model is $\frac{1}{Z} e^{-\beta E_k}$. Applying active inference and subsequent decision making to $\hat{\rho}$ entails the wavefunction collapse and converges to the ground eigenstate with the lowest eigenenergy, $E_0$, with high probability.

In the absence of feedback responses to the environment, the expectation of $\mathcal{A}$ is lower bounded by the Shannon entropy of the sensory inputs, $\langle \mathcal{A} \rangle_{p(\tilde{o})} \geq \langle -\log p(\tilde{o}) \rangle_{p(\tilde{o})}$, from Jensen's inequality. Consequently, parameter optimisation leads to the convergence of the Helmholtz energy to the input entropy. In cases wherein the variation for perturbations becomes zero, each component of $\mathcal{H}$ must correspond to that of $\mathcal{H}_E$, as shown in equation (10). Therefore, even in quantum systems, the system's Hamiltonian asymptotically converges to that of the environment.

This expectation was substantiated through the usage of the VQE (**Fig. 5a**). The VQE serves as a method for identifying the optimal solution by applying the variational principle $\langle\psi|\hat{H}|\psi\rangle \geq E_0$. Interestingly, when the VQE received the same sensory inputs as in the in vitro study (**Fig. 4**), the VQE combined with the conjugate parameter updating could perform blind source separation. The optimisation results—after sufficient iteration of updates of the state vector and parameters—confirmed the occurrence of Hamiltonian matching (**Fig. 5b**). These results underscore the applicability of Bayesian mechanics to make quantum computers infer the external milieu and to create quantum artificial intelligence.

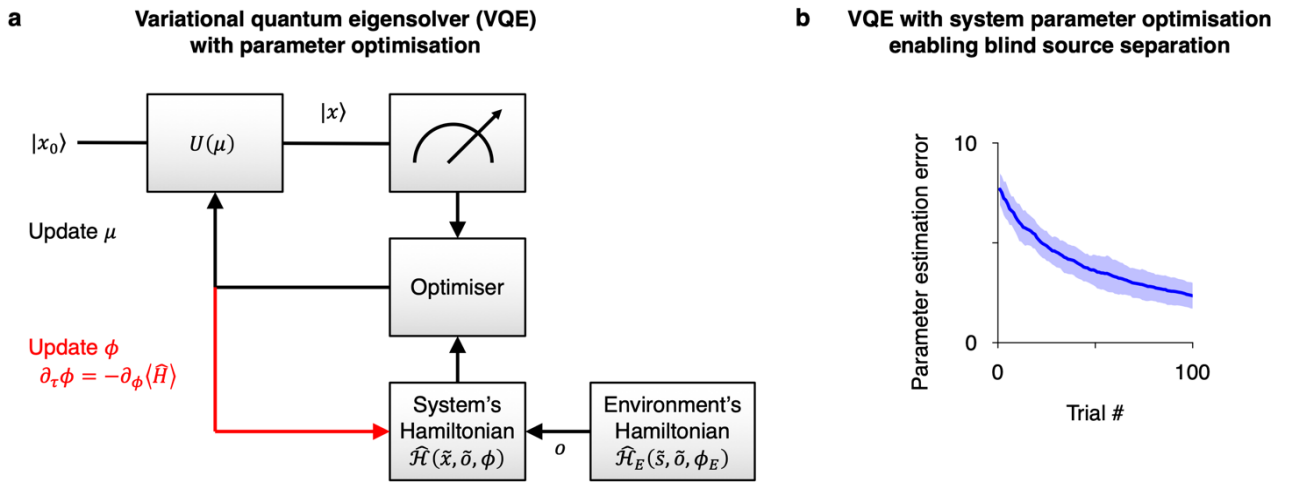

**Fig. 5. Bayesian mechanics in variational quantum eigensolvers. a,** A quantum system with *N* qubits $x \in \{0,1\}^N$ is expressed by a $2^N$-dimensional state vector $|x\rangle$. The VQE is a quantum algorithm for solving optimisation problems using the minimisation of the Hamiltonian [18]. It identifies an approximate optimal solution (ground states) of $\hat{H}$ using the variational principle

$\langle\psi|\widehat{H}|\psi\rangle \geq E_0$. The Hamiltonian is characterised by classical parameters $\phi$ and sensory inputs $o$; subsequently, the VQE renders a plausible hidden state for a given sensory input. The state vector $|x\rangle$ (the counterpart of the steady-state distribution $\pi$) is represented using a unitary operator $U(\mu)$ as $|x\rangle = U(\mu)|x_0\rangle$, where $U(\mu)$ is formed using quantum gates parameterised by sufficient statistics $\mu$. This construction of $|x\rangle$ using $\mu$ is formally homologous to an approximate posterior belief characterised by certain statistics. **b,** When the state vector and system parameters are adjusted to minimise the shared Hamiltonian, they effectively perform variational Bayesian inference of the hidden states, resulting in a recapitulation of the external system. The same sensory inputs as in **Fig. 4** were used. The errors in estimating parameters decreased with trials, consistent with the theoretical notion that the system's Hamiltonian becomes equivalent to that of the environment. In short, the extension of Bayesian mechanics from classical to quantum systems was demonstrated using VQEs by showing the convergence of the internal Hamiltonian with the external Hamiltonian led by minimising the shared Helmholtz energy.

**Remarks and future directions**

This work delved into Bayesian mechanics in realistic self-organising systems, encompassing classical, neural, and quantum systems. Reverse engineering is essential for underpinning Bayesian mechanics by identifying the implicit cost functions of dynamical systems, without directly solving the equations of dynamics. Subsequently, Bayesian mechanics facilitates the characterisation of their final (steady) states from the perspective of Bayesian inference, bypassing the requirement to directly compute time or path integrals. Although most practical dynamical systems do not have analytical solutions, through the lens of Bayesian inference, one can gain insight into the states in which these systems self-organise over long-term interactions with their external milieu. In essence, arbitrary self-organising systems that engage with external environments aim to recapitulate or synchronise with those external systems (i.e., Hamiltonian matching). These self-organising dynamics can be interpreted as a Bayesian agent operating under the hypothesis that the external milieu generates sensory inputs by following their own Hamiltonian. In addition, when the update of certain parameters is sufficiently slower than that of others, the system effectively aims to alter the environmental states to minimise $\Delta \mathcal{A}$, leading to active inference [22].

Quantum Bayesian mechanics postulates that quantum uncertainty is the origin of uncertainties and fluctuations within internal states, rather than attributing them to thermal noise. Quantum properties, including superposition, entanglement, and wavefunction collapse, may offer advantages over classical computation, especially in terms of parallel processing capabilities, and open avenues for (quasi-)quantum computation in neural systems. Particularly, if the interacting external milieu is a quantum system, quantum Bayesian mechanics would be useful for making appropriate inferences. However, it should be noted that the quantum system considered herein is idealised, which assumes that assimilating sensory inputs does not interfere with the internal quantum states. To rigorously handle quantum systems under observation, one must consider the Lindblad equation [50], which leads to a complexity that exceeds the scope of this work. This becomes essential to account for the interference with internal and external quantum states by observation and action, which will be an important direction for future work.

In summary, Bayesian mechanics enables the application of strong predictions of Bayesian inference to characterise generic dynamical systems; namely, internal dynamics self-organise to recapitulate the dynamics of the interacting external milieu. The demonstrations featuring in vitro neural networks and VQEs substantiated the potential applicability of this notion to various real-world systems. This notion will be crucial in understanding the mechanisms underlying the emergence of intelligence in biological organisms and creating biomimetic artificial intelligence.

**Methods**

***Sparse coupling***

In the Bayesian mechanics literature [6–10], it is commonly assumed that the system interacts with the environment through 'sparse coupling' to achieve sparse information flow at the steady state. This can be achieved when a dynamical system exhibits a unique attractor (eigenattractor) that does not directly depend on the sensory inputs, that is, $f$ is a function solely of $x$. This occurs in a wide range of dynamical systems interacting sparsely with their environment, as described below.

Here, sparse coupling is defined as a condition in which a solenoidal flow $f(x,o)$ receives a minor contribution from $o$. A substantial portion of $o$ can be well predicted using a function of $x$, denoted as $h = h(x)$, with a sufficiently small prediction error such that the second- or higher-order residual $\mathcal{O}(|o-h|^2)$ is negligible. In the Taylor expansion $f(x,o) = f(x,h) + \frac{\partial f}{\partial o}\Big|_{o=h}(o-h) + \mathcal{O}(|o-h|^2)$, the first-order term becomes zero, $\partial_x \cdot \frac{\partial f}{\partial o}\Big|_{o=h} = \frac{\partial}{\partial o}(\partial_x \cdot f)\Big|_{o=h} = 0$. This occurs because the solenoidal flow is orthogonal to the gradient flow, and thus $\partial_x \cdot f = 0$

holds by its construction. This implies that the contribution of $o$ can be involved in the scalar potential $V(x, o)$. Therefore, when $\mathcal{O}(|o - h|^2)$ is negligible, $f(x, o)$ can be reduced to a function solely of $x$, that is, $f(x) = f(x, h)$.

A small residual implies that $o$ provides only minor innovation of information, resulting in a minimal information flow from external milieu. In this context, sparse coupling indicates the sparseness of the information flow, i.e., the internal and external states are conditionally independent. This definition aligns with the definition involving the Markov blanket [6,9] and other relevant literature such as the Kalman filter [51]. Crucially, however, this remains well-defined even in non-steady states unlike the one using the Markov blanket.

### *Hamiltonian of the system*

This section elaborates that a broad range of dynamical systems can be formulated as a gradient descent on the Hamiltonian. In this work, an imaginary-time gradient descent of the path is introduced to address non-steady-state dynamics. Although this derivation introduces variations to the established body of Bayesian dynamics literature [6–10], the current work embraces this approach for its applicability in connecting Bayesian mechanics to generic physical systems.

This formulation is based on the generalised coordinates of motion [4,9]. Replacing $f$ in equation (1) by $x'$ introduces a minor delay $\Delta t$ ($0 < \Delta t \ll 1$) because $x'$ is calculated based on the information from $x$ and $o$, such as $x'(t + \Delta t) = f$, and their difference can be quantified as a quadratic error. The Taylor expansion $x'(t + \Delta t) = x' + \dot{x}'\Delta t + \mathcal{O}(\Delta t^2)$ is adopted for a small value of $\Delta t$. The two equalities imply the update rule $\dot{x}' \propto -(x' - f)$ that can be read as a gradient descent on the implicit kinetic energy or quadratic error $\frac{1}{2}|x' - f|^2$. Similarly, from $a(t + \Delta t) = g$, the dynamics of $a$ can be viewed as minimising $\frac{1}{2}|a - g|^2$. Moreover, $\dot{x}$ should match $x'$ for consistency, leading to an additional quadratic error $\frac{\Delta t}{2}|\dot{x} - x'|^2$. Hence, by combining these with the original potential $V$, the Hamiltonian $\mathcal{H}$ for the considered system is defined as equation (2).

The dynamics of $u = (x^T, x'^T, a^T)^T$ in the presence of force is expressed as equation (3). This can be associated with a particle interacting with the scalar potential $\mathcal{H}$ (**Fig. 1b**, left). If $\partial_u \mathcal{H}$ is explicitly written down, then equation (3) can be expressed as follows:

$$\begin{cases} \dot{x} = \frac{1}{\Delta t}(-\partial_x \mathcal{H} + \xi_x) \Leftrightarrow \dot{x} = \frac{1}{1 + \Delta t}\{x' + f'^T(x' - f) + g'^T(a - g) - \partial_x V + \xi_x\} \\ \dot{x}' = \frac{1}{\Delta t}(-\partial_{x'} \mathcal{H} + \xi_{x'}) = \frac{1}{\Delta t}\{-(x' - f) + \xi_{x'}\} \\ \dot{a} = \frac{1}{\Delta t}(-\partial_a \mathcal{H} + \xi_a) = \frac{1}{\Delta t}\{-(a - g) + \xi_a\} \end{cases} \quad (11)$$

where $f' = \partial_x f$ and $g' = \partial_x g$ are the Jacobi matrices of $f$ and $g$, respectively, and $\dot{x}' = \partial_t x'$ is the real-time derivative of $x'$. To compute the first equation, $\partial_x \dot{x} = \partial_{x(t)} \frac{x(t) - x(t - \Delta t)}{\Delta t} = \frac{1}{\Delta t}$ was adopted, analogous to the difference equation with sufficiently small $\Delta t$. Equation (11) follows conventional predictive coding models [4,9]. Transformation from equation (1) to equation (11) yields an uncertainty in the order of $\Delta t$. In the limit of small time constant $\Delta t \to 0$, equation (11) reduces to the original dynamical system described in equation (1), where $\xi_{x'} = 0$.

Bayesian mechanics operates under Euclidean space and the evolution of the path, where the Wick rotation [21] facilitates updating the path vector and parameters in imaginary time. The path calculation has the advantage of rendering the analysis considerably more manageable than that for the states at each time point that exhibit steady-state dynamics [8]. Within Euclidean space, the conventional Hamilton's canonical equations of motion are replaced by gradient descent on the Hamiltonian, in which the kinetic energies about $x'$ and $a$, $\frac{1}{2}|x'-f|^2 + \frac{1}{2}|a-g|^2$, are associated with elastic potentials.

Notably, the external (i.e., environmental) dynamics can be formulated in the same form as equation (1) by replacing the internal states $\varphi = \{\tilde{x}, \tilde{x}', \tilde{a}, \phi\}$ with the external states $\varphi_E = \{\tilde{s}, \tilde{s}', \tilde{o}, \theta\}$. With this replacement, the resulting environmental Hamiltonian $\mathcal{H}_E(\varphi_E, \tilde{a})$ is defined by substituting $(\varphi, \tilde{o}) = (\varphi_E, \tilde{a})$ into equation (2). This $\mathcal{H}_E$ forms the generative process of the external milieu, $p(\varphi_E, \tilde{a}) \propto e^{-\mathcal{H}_E(\varphi_E, \tilde{a})}$.

In generic dynamical systems, the action-reaction law does not necessarily hold; thus, the internal and external Hamiltonians can select in a mutually independent manner, resulting in equation (2). Alternatively, when the action-reaction law holds, the total Hamiltonian can be decomposed into the internal ($\mathcal{H}_I$), boundary ($\mathcal{H}_B$), and external ($\mathcal{H}_O$) Hamiltonians. As observation $o$ equates to action $a$, symmetry is required for the boundary or blanket-state Hamiltonian $\mathcal{H}_B(\tilde{a}, \tilde{o})$. Under a weak interaction between internal and external systems, it is given as $\mathcal{H}_B(\tilde{a}, \tilde{o}) = \int_0^t \frac{\varepsilon}{2\Delta t}|a-o|^2 dt'$. The internal Hamiltonian $\mathcal{H}_I(\varphi)$ is a function only of $\varphi$, $\mathcal{H}_I(\varphi) = \int_0^t \left\{ \frac{1}{2}|x' - f(x)|^2 + \frac{1}{2}|a - g(x)|^2 + \frac{\Delta t}{2}|\dot{x} - x'|^2 + V(x) \right\} dt' + C(\phi)$. The internal states follow the system's Hamiltonian, expressed as the sum of internal and boundary Hamiltonians, $\mathcal{H}(\varphi, \tilde{o}) = \mathcal{H}_I(\varphi) + \mathcal{H}_B(\tilde{a}, \tilde{o})$.

For instance, the dynamics of oscillators in **Fig. 3a, b** are characterised by the Stuart-Landau equation [23], $\dot{x} = x + \Omega x - |x|^2 x$, where $x = \begin{pmatrix} x_1 \\ x_2 \end{pmatrix}$ denotes two-dimensional states and $\Omega = \begin{pmatrix} 0 & -\omega \\ \omega & 0 \end{pmatrix}$ denotes parameter matrix with eigenfrequency $\omega$. In this case, the system's

Hamiltonian is reverse engineered as $\mathcal{H}(\varphi, \tilde{o}) = \int_0^t \left\{ \frac{1}{2}|x' - \Omega x|^2 + \frac{1}{2}|a - \Omega x|^2 + \frac{\varepsilon}{2}|a - o|^2 + \frac{\Delta t}{2}|\dot{x} - x'|^2 + V(x) \right\} dt' + C$ with a scalar potential $V(x) = -\frac{1}{2}|x|^2 + \frac{1}{4}|x|^4$.

### *Derivation of the non-steady-state Helmholtz energy*

Reverse engineering enables the derivation of the non-steady-state Helmholtz energy from the Fokker–Planck equation in a systematic and bottom-up manner. For preparation, first, the deterministic dynamics of a single particle are investigated. Here, $\partial_\tau \varphi = -\partial_\varphi \mathcal{H}$ is shown to be a reasonable update rule when the internal states $\varphi = \{\tilde{u}, \phi\}$ minimise the Hamiltonian $\mathcal{H}$ in the absence of noise. Consider $\partial_\tau \varphi = X(\varphi)$ as a suitable update rule. In the absence of any boundary, $|X|^2$ diverges unboundedly. To avoid this divergence, a constraint on the magnitude of $|X|^2$ is imposed. Although adopting a Riemannian metric yields a natural gradient [52], here, the Euclidean metric $|X|^2$ is adopted for simplicity. From the Lagrange multiplier method, the cost function is given as $\partial_\tau \mathcal{H} + \lambda |X|^2 = \partial_\varphi \mathcal{H} \cdot \partial_\tau \varphi + \lambda |X|^2 = \partial_\varphi \mathcal{H} \cdot X + \lambda |X|^2$. The solution that minimises this cost function is given by the condition that the variation is zero, $\delta(\partial_\varphi \mathcal{H} \cdot X + \lambda |X|^2) = (\partial_\varphi \mathcal{H} + 2\lambda X) \cdot \delta X = 0$. Hence, the solution of $X$ that minimises the cost function is $X \propto -\partial_\varphi \mathcal{H}$. Therefore, under the constraint on the magnitude of $|X|^2$, the optimal update rule that minimises $\partial_\tau \mathcal{H}$ with the highest efficiency is $\partial_\tau \varphi = -\partial_\varphi \mathcal{H}$, up to a constant multiplier. The amount of energy reduction is expressed as $\partial_\tau \mathcal{H} = \partial_\varphi \mathcal{H} \cdot \partial_\tau \varphi = -|\partial_\tau \varphi|^2 = -|\partial_\varphi \mathcal{H}|^2$.

Next, the aforementioned derivation is extended to a distribution to identify the plausible cost function in the presence of noise. The non-steady-state distribution of $\varphi$ is defined as $\pi(\varphi) = p(\varphi, \tau | \tilde{o})$. Using this, the Fokker-Planck equation corresponding to equation (4) is given as $\partial_\tau \pi =$ $-\partial_\varphi \cdot \left(\pi(-\partial_\varphi \mathcal{H})\right) + \frac{1}{\beta}\partial_\varphi \cdot \partial_\varphi \pi$, or, equivalently, equation (5). Subsequently, consider $\partial_\tau \pi = \partial_\varphi \cdot Y$ as the update rule that minimises a given cost function $\mathcal{A} = \int X\pi d\varphi = \langle X \rangle_\pi$. Because $\partial_\tau(X\pi) = X\partial_\tau \pi + \pi \partial_\tau X = \partial_\tau \pi (X + \pi \partial_\pi X)$, the imaginary-time derivative of $\mathcal{A}$ is given as follows:

$$\partial_\tau \mathcal{A} = \int \partial_\tau (X\pi) d\varphi = \int (\partial_\varphi \cdot Y)(X + \pi \partial_\pi X) d\varphi \tag{12}$$

From the integration by parts, $\partial_\tau \mathcal{A} = [(Y \cdot 1)(X + \pi \partial_\pi X)]_\Phi - \int Y \cdot \partial_\varphi (X + \pi \partial_\pi X) d\varphi = -\int Y \cdot \partial_\varphi (X + \pi \partial_\pi X) d\varphi$ holds, where $[(Y \cdot 1)(X + \pi \partial_\pi X)]_\Phi = 0$, as shown below. By the Lagrange multiplier method, the $Y$ that minimises $\partial_\tau \mathcal{A} + \lambda \langle |Y|^2 \rangle_\pi = \int \{-Y \cdot \partial_\varphi(X + \pi \partial_\pi X) + \lambda |Y|^2 \pi\} d\varphi$ is given as follows:

$$\delta \int \{-Y \cdot \partial_\varphi(X + \pi \partial_\pi X) + \lambda |Y|^2 \pi\} d\varphi = \int \{-\delta Y \cdot \partial_\varphi(X + \pi \partial_\pi X) + 2\lambda \pi Y \cdot \delta Y\} d\varphi$$

$$= \int \delta Y \cdot \{-\partial_\varphi(X + \pi \partial_\pi X) + 2\lambda \pi Y\} d\varphi = 0$$

$$\Leftrightarrow Y = \frac{1}{2\lambda \pi} \partial_\varphi (X + \pi \partial_\pi X) \tag{13}$$

When $X = \mathcal{H} + \frac{1}{\beta} \log \pi$ and $\lambda = \frac{1}{2\pi^2}$ hold, $Y = \pi \partial_\varphi \left(\mathcal{H} + \frac{1}{\beta} \log \pi\right)$ is obtained from equation (13). Therefore, this pair of $X$ and $Y$ satisfies both equations (5) and (6).

In essence, the cost function $\mathcal{A}$ that the Fokker–Planck equation (equation (5)) effectively minimises was reverse engineered. The implicit cost function was obtained in the form of the non-steady-state Helmholtz energy (equation (6)). This, in turn, provides a justification for the adoption of equation (6) as the cost function of the considered system. Solving for the fixed point of equation (5), which satisfies $\partial_\tau \pi = 0$, yields the steady-state distribution given by equation (7). Furthermore, equation (7) can be derived by solving $\pi(\varphi)$ that minimises equation (6) via the variational method $\delta \mathcal{A} = \int \left(\mathcal{H} + \frac{1}{\beta} \log \pi(\varphi)\right) \delta \pi(\varphi) d\varphi = 0$ under the $\int \pi(\varphi) d\varphi = 1$ constraint.

Notably, the identified Helmholtz energy $\mathcal{A}$ formally corresponds to the variational free energy $\mathcal{F}$ (equation (8)), owing to the one-to-one correspondences between components of the Hamiltonian and generative model (**Fig. 2**). That is, minimising $\mathcal{A}$ is naturally equivalent to performing variational Bayesian inference by minimising $\mathcal{F}$ under a specific generative model characterised by the system's Hamiltonian. Substituting equation (7) into equation (6) yields the steady-state Helmholtz energy $\mathcal{A}_s = -\frac{1}{\beta}\log Z$, which satisfies $\mathcal{A} \geq \mathcal{A}_s$ for any form of $\pi(\varphi)$. Similarly, substituting the posterior distribution (equation (9)) into equation (8) results in the surprisal $\mathcal{F}_s = -\log p_m(\tilde{o})$, which satisfies $\mathcal{F} \geq \mathcal{F}_s$ for any $q(\vartheta)$. This results in $\mathcal{A}_s \equiv \mathcal{F}_s$ and $Z \equiv p_m(\tilde{o})$ as long as the integration constant $C$ is set to satisfy $\int Z d\tilde{o} = 1$ in the steady state.

*Derivation of Hamiltonian matching*

In an adiabatic approximation, the steady-state distribution is expressed as $\pi(\varphi) = \pi(\tilde{u})\pi(\phi)$. To achieve Hamiltonian matching, $\pi(\phi)$ must be optimised with respect to the expectation over the distribution $p(\tilde{o})$. The information from previous sessions has been accumulated in $C(\phi)$ to generate the prior belief regarding the parameters, which is updated with information from the current session, providing the posterior belief $q(\theta) = \pi(\phi)$.

The distribution of the path of sensory inputs $p(\tilde{o})$ is determined by the generative process of the external milieu, where marginalisation yields $p(\tilde{o}) = \int p(\varphi_E, \tilde{a}) d\tilde{a} d(\varphi_E \backslash \tilde{o})$, where $\varphi_E \backslash \tilde{o} = \{\tilde{s}, \tilde{s}', \theta\}$ indicates a set of external states other than $\tilde{o}$. Thus, the input distribution is a functional of the environment's Hamiltonian $\mathcal{H}_E$, represented as $p(\tilde{o}) = \int e^{-\mathcal{H}_E(\varphi_E, \tilde{a})} d\tilde{a} d(\varphi_E \backslash \tilde{o})$. Based on the non-negativity of the Kullback–Leibler divergence, $\mathcal{D}[p(\tilde{o})||p_m(\tilde{o})] = \langle \log p(\tilde{o}) -$ $\log p_m(\tilde{o})\rangle_{p(\tilde{o})} \geq 0$ is satisfied. Thus, the expected value of surprisal, or equivalently steady-state Helmholtz energy, exceeds the input entropy: $\langle \mathcal{A}_s \rangle_{p(\tilde{o})} = \langle -\log p_m(\tilde{o}) \rangle_{p(\tilde{o})} \geq \langle -\log p(\tilde{o}) \rangle_{p(\tilde{o})}$. As the equality holds only when $\langle \mathcal{A}_s \rangle_{p(\tilde{o})}$ aligns with the input entropy $\langle -\log p(\tilde{o}) \rangle_{p(\tilde{o})}$, the self-organisation of internal states occurs to match $\mathcal{A}_s$ with $-\log p(\tilde{o})$; consequently, $\mathcal{A}_s = -\log p(\tilde{o})$ asymptotically holds after sufficient self-organisations.

To investigate the detail properties, the differences between the internal and external Helmholtz energies

$$\Delta \mathcal{A} = \langle \mathcal{A} - \mathcal{A}_E \rangle_{p(\varphi, \varphi_E)}$$

$$= \int \left( \mathcal{H}(\varphi, \tilde{o}) + \frac{1}{\beta} \log \pi(\varphi) - \mathcal{H}_E(\varphi_E, \tilde{a}) - \frac{1}{\beta} \log \pi_E(\varphi_E) \right) p(\varphi, \varphi_E) d\varphi d\varphi_E \quad (14)$$

is considered, where $p(\varphi, \varphi_E)$ denotes the joint probability of internal and external states. To make $\Delta \mathcal{A}$ retain zero against arbitrary perturbations, $\delta \Delta \mathcal{A} = 0$, the following condition needs to be satisfied under the steady state:

$$\delta \int \left( \mathcal{H} + \frac{1}{\beta} \log \pi - \mathcal{H}_E - \frac{1}{\beta} \log \pi_E \right) p(\varphi, \varphi_E) d\varphi d\varphi_E = 0$$

$$\Leftrightarrow \int \left\{ \delta(\mathcal{H} - \mathcal{H}_E) p + \left( \mathcal{H} + \frac{1}{\beta} \log \pi - \mathcal{H}_E - \frac{1}{\beta} \log \pi_E \right) \delta p \right\} d\varphi d\varphi_E = 0$$

$$\Leftrightarrow \int \left\{ \delta(\mathcal{H} - \mathcal{H}_E) p - \frac{1}{\beta} (\log Z - \log Z_E) \delta p \right\} d\varphi d\varphi_E = 0$$

$$\Leftrightarrow \int \delta(\mathcal{H} - \mathcal{H}_E) p d\varphi d\varphi_E = 0 \quad (15)$$

The transformation in the last line occurs because partition functions $Z$ and $Z_E$ are not functions of $\{\tilde{x}, \tilde{x}', \phi, \tilde{s}, \tilde{s}', \theta\}$, and thus their integrals over $\delta p(\varphi, \varphi_E)$ are zero. Therefore, the condition $\delta(\mathcal{H} - \mathcal{H}_E) = 0$ is required for the variation $\delta \Delta \mathcal{A}$ to be zero.

The variations in the Hamiltonians of the system and environment are expressed as

$$\begin{cases} \delta\mathcal{H} = \int_0^t (\partial_x\mathcal{H} \cdot \delta x + \partial_{x'}\mathcal{H} \cdot \delta x' + \partial_a\mathcal{H} \cdot \delta a + \partial_o\mathcal{H} \cdot \delta o)dt' + \partial_\phi\mathcal{H} \cdot \delta\phi \\ \delta\mathcal{H}_E = \int_0^t (\partial_s\mathcal{H}_E \cdot \delta s + \partial_{s'}\mathcal{H}_E \cdot \delta s' + \partial_a\mathcal{H}_E \cdot \delta a + \partial_o\mathcal{H}_E \cdot \delta o)dt' + \partial_\theta\mathcal{H}_E \cdot \delta\theta \end{cases} \quad (16)$$

Therefore, the conditions to be $\delta(\mathcal{H} - \mathcal{H}_E) = 0$ are explicated as follows:

$$\begin{cases} \partial_x\mathcal{H} \cdot \delta x = \partial_s\mathcal{H}_E \cdot \delta s \\ \partial_{x'}\mathcal{H} \cdot \delta x' = \partial_{s'}\mathcal{H}_E \cdot \delta s' \\ \partial_a\mathcal{H} \cdot \delta a = \partial_a\mathcal{H}_E \cdot \delta a \\ \partial_o\mathcal{H} \cdot \delta o = \partial_o\mathcal{H}_E \cdot \delta o \\ \partial_\phi\mathcal{H} \cdot \delta\phi = \partial_\theta\mathcal{H}_E \cdot \delta\theta \end{cases} \quad (17)$$

This is referred to as Hamiltonian matching. Because $o$ and $a$ are shared between the internal system and the external environment, $\partial_o\mathcal{H} = \partial_o\mathcal{H}_E$ and $\partial_a\mathcal{H} = \partial_a\mathcal{H}_E$ are required to retain $\delta(\mathcal{H} - \mathcal{H}_E) = 0$ in the presence of perturbations. Moreover, the ratio of variations, $\delta s_i/\delta x_j$ and $\delta s'_i/\delta x'_j$, can be associated with the Jacobi matrices $\partial_x s$ and $\partial_{x'} s'$, respectively. Therefore, equation (10) is derived.

*Derivation of STDP from a realistic neural activity model*

The Hodgkin–Huxley model [27] comprises the four differential equations, in which the first equation is for the electrical circuit that generates spike activity, whereas the other three express the dynamics of the variables controlling the magnitude of the non-linear variable conductance. This is given as follows:

$$\begin{cases} C_M\dot{v} = -g_{Na}m^3h(v - E_{Na}) - g_Kn^4(v - E_K) - g_L(v - E_L) + I \\ \dot{m} = \alpha_m(v)(1 - m) - \beta_m(v)m \\ \dot{h} = \alpha_h(v)(1 - h) - \beta_h(v)h \\ \dot{n} = \alpha_n(v)(1 - n) - \beta_n(v)n \end{cases} \quad (18)$$

where $v$ denotes the dynamics of membrane potential; $(m, h)$ and $n$ denote the gating variables for sodium and potassium ion channels, respectively; $C_M$ denotes a membrane capacitance; $(E_{Na}, E_K, E_L)$ and $(g_{Na}, g_K, g_L)$ denote the reversal potential and conductance for sodium, potassium, and leak channels, respectively; functions $\alpha_i(v)$ and $\beta_i(v)$ for $i = m, h, n$ characterise the dynamics of gating variables; and $I$ denotes the external input. Subsequently, the spike activity is transmitted to the postsynaptic (i.e., subsequent) neuron by releasing neurotransmitters such as glutamic acid. Glutamatergic synapses express α-amino-3-hydroxyl-5-methyl-4-isoxazole-propionate (AMPA) and *N*-methyl-D-aspartate (NMDA) receptors. In this case, $I$ in equation (18) denotes the post synaptic current generated from the AMPA and NMDP receptor dynamics [53]:

$$I = -g_{AMPA}(v - 0) - g_{NMDA} \frac{\left(\frac{v+80}{60}\right)^2}{1 + \left(\frac{v+80}{60}\right)^2}(v - 0) \quad (19)$$

The conductance of AMPA and NMDA receptors can be modelled as $g_{AMPA} = W(\eta_{AMPA} * o)$ and $g_{NMDA} = W(\eta_{NMDA} * o)$ using synaptic weight $W$ and filtered input $\eta * o$, where $\eta *$ denotes convolution with exponential filter such as $\eta_{AMPA} = e^{-t/t_{AMPA}}\Theta(t)$ and $\eta_{NMDA} = e^{-t/t_{NMDA}}\Theta(t)$, where $t_{AMPA}$ = 5 ms, $t_{NMDA}$ = 150 ms, and $\Theta(t) = 0$ for $t \leq 0$ and 1 for $t > 0$.

From the perspective of Bayesian mechanics, both neural activity ($u$) and synaptic weights ($W \in \phi$) commonly minimise the shared Hamiltonian. Equations (18) and (19) can be viewed as a family of equation (1) when $x = (v, m, h, n)^T$. The membrane dynamics are expressed by the solenoidal flow $f = (f_v, f_m, f_h, f_n)^T$, in which $f_v = \frac{1}{C_M}\{-g_{Na}m^3h(v - E_{Na}) - g_K n^4(v - E_K) - g_L(v - E_L)\}$ and $f_i = \alpha_i(v)(1 - i) - \beta_i(v)i$ for $i = m, h, n$. Moreover, the post synaptic current is expressed by the gradient flow, $I = -\partial_v V$. The scholar potential is computed as $V = W\{\frac{1}{2}(v^2 - $ $v_r^2)(\eta_{\text{AMPA}} * o) + \big(J(v) - J(v_r)\big)(\eta_{\text{NMDA}} * o)\big\}$, where $J(v) = \frac{(v+80)^2}{2} - \frac{60^2}{2}\ln\left(1 + \left(\frac{v+80}{60}\right)^2\right) - 60 \cdot 80\left(\frac{v+80}{60} - \tan^{-1}\left(\frac{v+80}{60}\right)\right)$ and $v_r$ denotes a constant that represents the resting membrane potential. By substituting these components into equation (2), the Hamiltonian $\mathcal{H}$ for the Hodgkin–Huxley model can be reverse engineered. Subsequently, the gradient descent on $\mathcal{H}$ with respect to $W$ (i.e., $\dot{W} = -\partial_W \mathcal{H}$) yields the following synaptic plasticity rule:

$$\dot{W} = \frac{1}{2}(v_r^2 - v^2)(\eta_{\text{AMPA}} * o)^{\text{T}} + \big(J(v_r) - J(v)\big)(\eta_{\text{NMDA}} * o)^{\text{T}} \tag{20}$$

As shown in **Fig. 2d**, the time window of equation (20) exhibits the empirically observed form of STDP [29].

### *Bayesian mechanics in VQEs*

By adopting the canonical quantisation of the Hamiltonian, $\widehat{\mathcal{H}}$, one arrives at the time-independent Schrödinger equation, $\widehat{\mathcal{H}}|\psi_k\rangle = E_k|\psi_k\rangle$, where $|\psi_k\rangle$ and $E_k$ represent eigenstates (eigenvectors) that constitute an orthonormal basis and eigenenergies (eigenvalues), respectively. Under the thermal equilibrium, the density matrix of the mixed state is provided as $\hat{\rho} = \frac{1}{Z}\sum_k e^{-\beta E_k}|\psi_k\rangle\langle\psi_k|$, where $Z = \sum_k e^{-\beta E_k}$ is the partition function. The steady-state Helmholtz energy is expressed as $\mathcal{A} = \text{tr}\left[\hat{\rho}\left(\widehat{\mathcal{H}} + \frac{1}{\beta}\log\hat{\rho}\right)\right]$, where $-\text{tr}[\hat{\rho}\log\hat{\rho}]$ denotes the von Neumann entropy. The variation of $\mathcal{A}$, $\delta\mathcal{A} = \text{tr}\left[\delta\hat{\rho}\left(\widehat{\mathcal{H}} + \frac{1}{\beta}\log\hat{\rho}\right)\right] = 0$, yields the thermal equilibrium density matrix as $\hat{\rho} = \frac{1}{Z}e^{-\beta\widehat{\mathcal{H}}}$. Thus, $\mathcal{A}$ can be transformed as $\mathcal{A} = -\frac{1}{\beta}\log Z$.

The VQE solves optimisation problems using the minimisation of the Hamiltonian [18]: $\widehat{\mathcal{H}}(\sigma) = -\sum_{i,j} J_{ij} Z_i Z_j - \sum_i h_i X_i$, where $Z_i$ indicates the Z-gate over the *i*-th element: $Z_i = I \otimes \cdots \otimes I \otimes$ $\overset{i}{\widetilde{Z}} \otimes I \otimes \cdots \otimes I$ and $X_i$ is the X-gate. The Z-gate takes eigenvectors such as $|0\rangle$ or $|1\rangle$, whereas the X-gate is an eigenvector of the superposition of the states. By minimising $\widehat{\mathcal{H}}(\sigma)$, one obtains the eigenvalues $E_i(\sigma)$ and the eigenvectors $|\psi_i(\sigma)\rangle$ that satisfy $\widehat{\mathcal{H}}(\sigma)|\psi_i(\sigma)\rangle = E_i(\sigma)|\psi_i(\sigma)\rangle$. Based on these, the configuration of $\sigma$ having $E_i(\sigma)$ at the inverse temperature $\beta$ follows the probability $p_i(\sigma) = \frac{1}{Z} e^{-\beta E_i(\sigma)}$, where the distribution function is $Z = \sum_i e^{-\beta E_i(\sigma)}$.

When Bayesian mechanics is applied to quantum systems, the internal state (of an agent) described as a state vector $|\psi_n\rangle$ can be viewed as the posterior distribution about the external states. For instance, $|x_{1:t}\rangle$ corresponds to the posterior expectation of the path of the hidden state of the POMDP $\mathbf{s}_{1:t}$, where $N$ elements can be expressed using a $2^N$ dimensional state vector. In this context, the density matrix of the mixed states $\hat{\rho}$ corresponds to the mixed generative model [40].

**Data Availability**

All relevant data are presented in this paper. Figs. 3–5 were generated using the author's scripts (see Code Availability).

**Code Availability**

The MATLAB scripts are available at https://github.com/takuyaisomura/reverse_engineering. The scripts are covered under the GNU General Public Licence v3.0.

**References**

1. Helmholtz, H. Treatise on Physiological Optics, Vol. 3. Optical Society of America, Washington, DC (1925).

2. Dayan, P., Hinton, G.E., Neal, R.M. & Zemel, R.S. The Helmholtz machine. *Neural Comput.* **7**, 889–904 (1995).

3. George, D. & Hawkins, J. Towards a mathematical theory of cortical micro-circuits. *PLoS Comput. Biol.* **5**, e1000532 (2009).

4. Friston, K. J., Kilner, J. & Harrison, L. A free energy principle for the brain. *J. Physiol. Paris* **100**, 70–87 (2006).

5. Friston, K. J. The free-energy principle: a unified brain theory? *Nat. Rev. Neurosci.* **11**, 127–138 (2010).

6. Friston, K. J. A free energy principle for a particular physics. arXiv preprint:1906.10184 (2019).

7. Da Costa, L., Friston, K., Heins, C. & Pavliotis, G. A. Bayesian mechanics for stationary processes. *Proc. R. Soc. A* **477**, 20210518 (2021).

8. Ramstead, M. J., Sakthivadivel, D. A., Heins, C. et al. On Bayesian mechanics: a physics of and by beliefs. *Interface Focus* **13**, 20220029. (2023).

9. Friston, K. J., Da Costa, L., Sajid, N. et al. The free energy principle made simpler but not too simple. *Phys. Rep.* **1024**, 1–29 (2023).

10. Friston, K. J., Da Costa, L., Sakthivadivel, D. A., Heins, C., Pavliotis, G. A., Ramstead, M., & Parr, T. Path integrals, particular kinds, and strange things. *Phys. Life Rev.* **47**, 35–62 (2023).

11. Wald, A. An essentially complete class of admissible decision functions. *Ann. Math. Stat.* **18**, 549–555 (1947).

12. Brown, L. D. A complete class theorem for statistical problems with finite-sample spaces. *Ann. Stat.* **9**, 1289–1300 (1981).

13. Berger, J. O. *Statistical Decision Theory and Bayesian Analysis* (Springer Science & Business Media, 2013).

14. Isomura, T. & Friston, K. J. Reverse-engineering neural networks to characterize their cost functions. *Neural Comput.* **32**, 2085–2121 (2020).

15. Isomura, T., Shimazaki, H. & Friston, K. J. Canonical neural networks perform active inference. *Commun. Biol.* **5**, 55, (2022).

16. Isomura, T. Active inference leads to Bayesian neurophysiology. *Neurosci. Res.* **175**, 38–45 (2022).

17. Isomura, T., Kotani, K., Jimbo, Y. & Friston, K. J. Experimental validation of the free-energy principle with in vitro neural networks. *Nat. Commun.* **14**, 4547 (2023).

18. Peruzzo, A., McClean, J., Shadbolt, P. et al. A variational eigenvalue solver on a photonic quantum processor. *Nat. Commun.* **5**, 4213 (2014).

19. Wick, G. C. Properties of Bethe-Salpeter wave functions. *Phys. Rev.* **96**, 1124–1134 (1954).

20. Friston, K. J., FitzGerald, T., Rigoli, F., Schwartenbeck, P. & Pezzulo, G. Active inference: A process theory. *Neural Comput.* **29**, 1–49 (2017).

21. Risken, H. & Frank, T. *The Fokker-Planck Equation: Methods of Solution and Applications, second ed.,* In: Springer Series in Synergetics, Springer-Verlag, Berlin Heidelberg (1996).

22. Esposito, M. & Van den Broeck, C. Second law and Landauer principle far from equilibrium. *Europhys. Lett.* **95**, 40004 (2011).

23. Stuart, J. T. On the non-linear mechanics of wave disturbances in stable and unstable parallel flows Part 1. The basic behaviour in plane Poiseuille flow. *J. Fluid Mech.* **9**, 353–370 (1960).

24. Blei, D. M., Kucukelbir, A. & McAuliffe, J. D. Variational inference: A review for statisticians. *J. Am. Stat. Assoc.* **112**, 859–877 (2017).

25. Mac Lane, S. Categories for the working mathematician (Vol. 5). Springer Science & Business Media (1998).

26. Friston, K. J. & Frith, C. D. Active inference, communication and hermeneutics. *Cortex* **68**, 129–143 (2015).

27. Hodgkin, A. L. & Huxley, A. F. A quantitative description of membrane current and its application to conduction and excitation in nerve. *J. Physiol.* **117**, 500–544 (1952).

28. Markram, H., Lübke, J., Frotscher, M. & Sakmann, B. Regulation of synaptic efficacy by coincidence of postsynaptic APs and EPSPs. *Science* **275**, 213–215 (1997).

29. Bi, G. Q. & Poo, M. M. Synaptic modifications in cultured hippocampal neurons: dependence on spike timing, synaptic strength, and postsynaptic cell type. *J. Neurosci.* **18**, 10464–10472 (1998).

30. Song, S., Miller, K. D. & Abbott, L. F. Competitive Hebbian learning through spike-timing-dependent synaptic plasticity. *Nat. Neurosci.* **3**, 919–926 (2000).

31. Song, Y., Millidge, B., Salvatori, T., Lukasiewicz, T., Xu, Z. & Bogacz, R. Inferring neural activity before plasticity as a foundation for learning beyond backpropagation. *Nat. Neurosci.* **27**, 348–358 (2024).

32. Rao, R. P. & Ballard, D. H. Predictive coding in the visual cortex: a functional interpretation of some extra-classical receptive-field effects. *Nat. Neurosci.* **2**, 79–87 (1999).

33. Friston, K. J. A theory of cortical responses. *Philos. Trans. R. Soc. Lond. B Biol. Sci.* **360**, 815–836 (2005).

34. Bastos, A. M., Usrey, W. M., Adams, R. A., Mangun, G. R., Fries, P. & Friston, K. J. Canonical microcircuits for predictive coding. *Neuron* **76**, 695–711 (2012).

35. Wolpert, D. M. & Kawato, M. Multiple paired forward and inverse models for motor control. *Neural Netw.* **11**, 1317–1329 (1998).

36. Gerstner, W., Kistler, W. M., Naud, R. & Paninski, L. *Neuronal Dynamics: From Single Neurons to Networks and Models of Cognition.* Cambridge University Press (2014).

37. FitzHugh, R. Impulses and physiological states in theoretical models of nerve membrane. *Biophys. J.* **1**, 445–466 (1961).

38. Nagumo, J., Arimoto, S. & Yoshizawa, S. An active pulse transmission line simulating nerve axon. *Proc. IRE* **50**, 2061–2070 (1962).

39. Hebb, D. O. *The Organization of Behavior: A Neuropsychological Theory* (Wiley, New York, 1949).

40. Isomura, T., Parr, T. & Friston, K. J. Bayesian filtering with multiple internal models: toward a theory of social intelligence. *Neural Comput.* **31**, 2390–2431 (2019).

41. Palacios, E. R., Isomura, T., Parr, T. & Friston, K. The emergence of synchrony in networks of mutually inferring neurons. *Sci. Rep.* **9**, 6412 (2019).

42. Vuust, P., Heggli, O. A., Friston, K. J. & Kringelbach, M. L. Music in the brain. *Nat. Rev. Neurosci.* **23**, 287–305 (2022).

43. Ahmadi, A. & Tani, J. A novel predictive-coding-inspired variational RNN model for online prediction and recognition. *Neural Comput.* **31**, 2025–2074 (2019).

44. Ito, H., Yamamoto, K., Mori, H. & Ogata, T. Efficient multitask learning with an embodied predictive model for door opening and entry with whole-body control. *Sci. Robot.* **7**, eaax8177 (2022).

45. Funamizu, A., Kuhn, B. & Doya, K. Neural substrate of dynamic Bayesian inference in the cerebral cortex. *Nat. Neurosci.* **19**, 1682–1689 (2016).

46. Berkes, P., Orbán, G., Lengyel, M. & Fiser, J. Spontaneous cortical activity reveals hallmarks of an optimal internal model of the environment. *Science* **331**, 83–87 (2011).

47. Torigoe, M. et al. Zebrafish capable of generating future state prediction error show improved active avoidance behavior in virtual reality. *Nat. Commun.* **12**, 5712 (2021).

48. Isomura, T., Kotani, K. & Jimbo, Y. Cultured cortical neurons can perform blind source separation according to the free-energy principle. *PLoS Comput. Biol.* **11**, e1004643 (2015).

49. Isomura, T. & Friston, K. J. In vitro neural networks minimise variational free energy. *Sci. Rep.* **8**, 16926 (2018).

50. Lindblad, G. On the generators of quantum dynamical semigroups. *Commun. Math. Phys.* **48**, 119–130 (1976).

51. Masreliez, C. & Martin, R. Robust Bayesian estimation for the linear model and robustifying the Kalman filter. *IEEE Trans. Autom. Control* **22**, 361–371 (1977).

52. Amari, S. I. Natural gradient works efficiently in learning. *Neural Comput.* **10**, 251–276 (1998).

53. Izhikevich, E. M., Gally, J. A. & Edelman, G. M. Spike-timing dynamics of neuronal groups. *Cereb. Cortex* **14**, 933–944 (2004).

**Acknowledgements**

T.I. is supported by the Japan Society for the Promotion of Science (JSPS) KAKENHI Grant Numbers JP23H04973 and JP23H03465 and the Japan Science and Technology Agency (JST) CREST Grant Number JPMJCR22P1. The funders had no role in study design, data collection and analysis, decision to publish, or preparation of the manuscript.

**Competing interest declaration**